\documentclass[prb, 11pt,superscriptaddress,floatfix,nofootinbib,notitlepage,longbibliography]{revtex4-1}
\usepackage{graphicx}
\usepackage[colorlinks=true,citecolor=blue]{hyperref}
\usepackage{amsmath}
\usepackage{amssymb}
\usepackage{soul}
\usepackage{gensymb}
\usepackage{xcolor}
\newcommand{\dIdV}{d$I$/d$V$}

\begin{document}

\title{Strain-induced two-dimensional topological crystalline insulator}

\author{Liwei Jing }
\affiliation{Department of Physics, Nanoscience Center, 
University of Jyväskyl\"a, FI-40014 University of Jyväskyl\"a, Finland}

\author{Mohammad Amini}
\thanks{These authors contributed equally to this work.}
\affiliation{Department of Applied Physics, Aalto University, FI-00076 Aalto, Finland}

\author{Adolfo O. Fumega}
\thanks{These authors contributed equally to this work.}
\affiliation{Department of Applied Physics, Aalto University, FI-00076 Aalto, Finland}

\author{Orlando J. Silveira}
\thanks{These authors contributed equally to this work.}
\affiliation{Department of Physics, Nanoscience Center, 
University of Jyväskyl\"a, FI-40014 University of Jyväskyl\"a, Finland}
\affiliation{Department of Applied Physics, Aalto University, FI-00076 Aalto, Finland}

\author{Jose L. Lado}
\email{e-mail: kezilebieke.a.shawulienu@jyu.fi; jose.lado@aalto.fi; peter.liljeroth@aalto.fi}
\affiliation{Department of Applied Physics, Aalto University, FI-00076 Aalto, Finland}

\author{Peter Liljeroth}
\email{e-mail: kezilebieke.a.shawulienu@jyu.fi; jose.lado@aalto.fi; peter.liljeroth@aalto.fi}
\affiliation{Department of Applied Physics, Aalto University, FI-00076 Aalto, Finland}

\author{Shawulienu Kezilebieke}
\email{e-mail: kezilebieke.a.shawulienu@jyu.fi; jose.lado@aalto.fi; peter.liljeroth@aalto.fi}
\affiliation{Department of Physics, Nanoscience Center, 
University of Jyväskyl\"a, FI-40014 University of Jyväskyl\"a, Finland}
\affiliation{Department of Chemistry, 
University of Jyväskyl\"a, FI-40014 University of Jyväskyl\"a, Finland}

\begin{abstract}

Topological crystalline insulators (TCIs) host topological phases of matter protected by crystal symmetries. Topological surface states in three-dimensional TCIs have been predicted and observed in IV-VI SnTe-class semiconductors. Despite the prediction of a two-dimensional (2D) TCI characterized by two pairs of edge states inside the bulk gap, materials challenges have thus far prevented its experimental realization. Here we report the growth and characterization of bilayer SnTe on the 2$H$-NbSe$_2$ substrate by molecular beam epitaxy and scanning tunneling microscopy. We experimentally observe two anticorrelated, periodically modulated pairs of conducting edge states along the perimeters of the sample with a large band gap exceeding $0.2$ eV. We identify these states with a 2D TCI through first principles calculations. Finally, we probe the coupling of adjacent topological edge states and demonstrate the resulting energy shift driven by a combination of electrostatic interactions and tunneling coupling. Our work opens the door to investigations of tunable topological states in 2D TCIs, of potential impact for spintronics and nanoelectronics applications at room temperature.

\end{abstract}

\date{\today}

\maketitle
\newpage

\section{Introduction}
Topological insulators (TIs) feature an energy gap in the bulk and nontrivial surface or edge states that connect the valence and conduction bands\cite{RevModPhys.82.3045,RevModPhys.83.1057,Moore2010}. The discovery of quantum spin Hall (QSH) effect in HgTe/CdTe quantum wells\cite{doi:10.1126/science.1148047} has sparked intensive research on novel topological materials extending from three-dimensional (3D) to two-dimensional (2D) systems\cite{Zhang2009,doi:10.1126/science.1173034,Hsieh2009,Drozdov2014,PhysRevX.6.021017,doi:10.1126/science.1256815}. Meanwhile, crystal symmetries are discovered to play an analogous role as time-reversal symmetry of QSH insulators, which can protect the boundary states from backscattering and thus result in dissipationless carrier transport\cite{Ando2015}. Unlike QSH described by $Z_2$ invariant, the topology in topological crystalline insulators (TCIs) is classified by the finite Chern number in each symmetry sector, having a net zero total Chern number. 3D TCIs have been realized in semiconductors tin telluride (SnTe)\cite{Hsieh2012,Tanaka2012}, Pb$_{1-x}$Sn$_x$Se\cite{doi:10.1126/science.1239451,Zeljkovic2014}, and Pb$_{1-x}$Sn$_x$Te\cite{Xu2012,ref1}, in which topological surface states are highly dependent on the surface orientations depending on whether time-reversal-mirror symmetry is broken or preserved. If the the time-reversal-mirror symmetry with respect to the (110) bulk mirror plane is broken, the Dirac cones would acquire a band gap and Dirac fermions would obtain mass\cite{doi:10.1126/science.1239451,Zeljkovic2015a}. The (001) SnTe films are predicted to be 2D TCIs both in the single atomic layer (AL) limit\cite{Liu2015} and above 5 ALs\cite{Liu2014}. They should host two pairs of edge states protected by the time-reversal-mirror symmetry $z\rightarrow-z$ ($z$ is normal to the film) through a nonzero time-reversal-mirror Chern number.

The hallmark of a 2D TCI is the presence of metallic edge states along the sample boundary. Higher-order one-dimensional (1D) topological hinge states\cite{Schindler2018,PhysRevLett.119.246401} and surface step edge states\cite{doi:10.1126/science.aah6233} have been theoretically predicted and experimentally observed, but it remains a challenge to realize a strong 2D TCI with a wide band gap due to the complicated fabrication of thin films and strong substrate-induced hybridization effects. Particularly, only the odd number of ALs are expected to be 2D TCIs because they keep the symmorphic crystal symmetry with non-zero time-reversal-mirror Chern number\cite{Hsieh2012,Araujo2018}. However, each monolayer of (001) SnTe ultrathin films in the rock-salt structure contains 2 ALs and thus we always get an even number of ALs in experiments\cite{Chang2019}, which prevents the band inversion necessary for a topological phase transition. Besides, a strong ferroelectric phase also emerges in the ultrathin limit\cite{doi:10.1126/science.aad8609,https://doi.org/10.1002/adma.202206456}. Realization of a 2D TCI could be achieved with strain, which allows tailoring the electronic band gap and electronic phase transitions of few-AL SnTe films\cite{Qian2015}. 

In this work, we utilize epitaxial growth to fabricate bilayer SnTe (4 ALs) islands on the bulk 2$H$-NbSe$_2$ and investigate atomic-scale structure and electronic properties using low-temperature scanning tunneling microscopy (STM) and spectroscopy (STS). We experimentally observe strong edge modes that can be rationalized through density functional theory (DFT) calculations showing the top 3 ALs of SnTe are decoupled from the first 1 AL due to strong interfacial interactions. The concomitant strain drives our effective 3-AL system to undergo a crossover from trivial ferroelectricity to a TCI phase. Finally, we probe experimentally the coupling of edge states from adjacent bilayer SnTe islands and reveal significant energy shifts. Our study highlights a pathway for creating strain-induced emergent quantum states of matter, which could significantly impact energy-efficient electronics.
\section{Results}

\subsection{\textbf{Experimental realization of strained 3 atomic layers}}

\begin{figure}[t!]
    \centering
    \includegraphics[width = 0.9\textwidth]{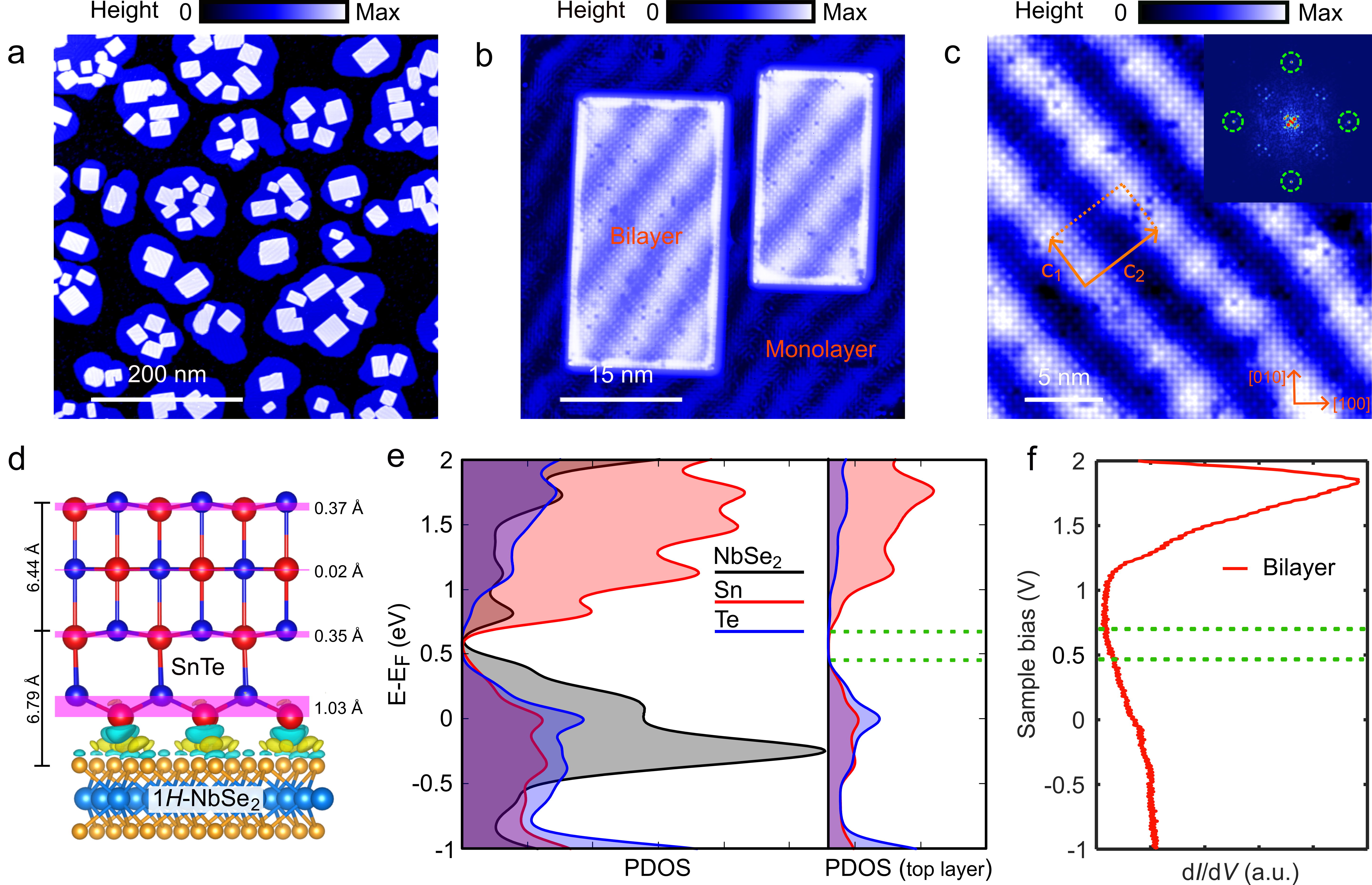}
    \caption{\textbf{Heteroepitaxial bilayer SnTe on 2$H$-NbSe$_2$.} \textbf{a}, Large-scale STM topographic image of bilayer SnTe on the 2$H$-NbSe$_2$ substrate (tunneling parameters: \textit{V$_b$}=1 V, \textit{I}=2.5 pA). \textbf{b}, Typical STM image of bilayer SnTe islands with visible stripe patterns(\textit{V$_b$}=0.8 V, \textit{I}=200 pA). \textbf{c}, Atomic-resolution STM topography of bilayer SnTe islands  (\textit{V$_b$}=0.3 V, \textit{I}=230 pA). Sublattice Sn atoms are aligned along [100] and [010] directions indicated by orange color, which corresponds to a set of Bragg lattice peaks outlined by green dashed circles in the inset. One of strain superlattices is marked by a nearly rectangular frame along {c$_1$} and {c$_2$} directions. \textbf{d}, Side view of the heterostructure model of 4-AL SnTe on 1$H$-NbSe$_2$ after full relaxation. The charge difference at the interface shown by green and yellow colors reveals that the bottom 1-AL SnTe bonds to the 1$H$-NbSe$_2$ underneath through Sn atoms, giving rise to its substantial decoupling from the top 3 ALs. The red, dark blue, orange, and light blue balls represent Sn, Te, Se and Nb atoms, respectively. \textbf{e}, Calculated PDOS of bilayer SnTe/1$H$-NbSe$_2$ by DFT methods. The left panel shows the PDOS components of all the Sn, Te and NbSe$_2$ orbitals onto themselves while the right panel displays PDOS of all the Sn and Te orbitals onto the Sn and Te atoms of the topmost atomic layer. \textbf{f}, Tunneling differential spectrum taken in the middle of bilayer SnTe. Two pairs of horizontal dashed green lines in \textbf{e} and \textbf{f} indicate the calculated and experimental bulk band gap edges of bilayer SnTe. }
\label{fig1}
\end{figure}

We prepared ultrathin SnTe films on a 2$H$-NbSe$_2$ substrate using the molecular beam epitaxy (MBE, see the Methods section for details on sample preparation). Since each monolayer consists of 2 ALs in the rock-salt SnTe structure, the growing island thickness is always an integer multiple of monolayer in ultrathin films. By carefully controlling the growth conditions, we achieved monolayer-by-monolayer growth. As shown in Figure \ref{fig1}a, the monolayer forms large amorphous islands where the bilayer subsequently nucleates. The height profile (see SI Figure S6) confirms that both layers are one monolayer thick. The stripes visible in Fig. \ref{fig1}b extend from the monolayer to the bilayer and they arise from the moiré patterns resulting from large mismatch between the square SnTe lattice (Te-Te distance of 4.5 \AA) and the hexagonal 2$H$-NbSe$_2$ lattice (Se-Se distance of 3.4 \AA). Monolayer SnTe islands exhibit a disordered structure with varying stripe orientations, serving as the wetting layer for the bilayer. In addition, the monolayer stripe patterns reflect a periodically varying strain field in the bilayer (see Extended Data in SI Figure S7). The strain field predominantly distributes along strained moiré patterns, with one unit cell of strain field being shown by orange color in Fig. \ref{fig1}c. By fast Fourier transform (in the inset) of atom-resolved bilayer SnTe topography in Fig. \ref{fig1}c, we find the lattice constants along the [100] and [010] directions are approximately 4.17 \AA and 4.33 \AA, respectively, which indicates bilayer SnTe is biaxially compressed. This scenario is quite different from the stress-free ultrathin SnTe films grown on graphene and graphite \cite{doi:10.1126/science.aad8609,https://doi.org/10.1002/adma.202206456}. 

To understand the 4-AL SnTe/2$H$-NbSe$_2$ heterostructure, we performed DFT calculations. Fig. \ref{fig1}d presents the side view of our heterostructure model after full relaxation (more details in SI Figure S8). The results suggest that strong hybridization between SnTe and 1$H$-NbSe$_2$ causes Sn atoms from the bottom 1 AL to be bonded to the substrate. The lattice mismatch further forces the top 3 ALs to be significantly decoupled from the bottom 1 AL with a substantial structural distortion. The interlayer height obtained from the DFT calculations is in good agreement with our experimental data. The projected density of states (PDOS) onto the top 1-AL SnTe is shown in Fig. \ref{fig1}e. The calculated PDOS agrees well with the experimentally measured tunneling differential spectrum (Fig. \ref{fig1}f), where the bulk band gap edges are marked by green dashed lines. Besides charge transfer with the substrate, Sn vacancies in bilayer SnTe further enhance its pronounced p-type character, lowering the Fermi level below the bulk valence band maximum\cite{PhysRevB.89.045142}. The estimated bilayer bulk band gap ranges between $0.2$ and $0.3$ eV, depending on the island size. A detailed analysis of the bulk gap can be found in the SI Figure S14. This analysis shows that the experiment effectively realizes a strained 3-AL SnTe system, with a 1-AL wetting layer on the 2$H$-NbSe$_2$ substrate.

\subsection{\textbf{Lifshitz transition driven by biaxial strain} }

\begin{figure}[t!]
    \centering
    \includegraphics[width = 0.9\textwidth]{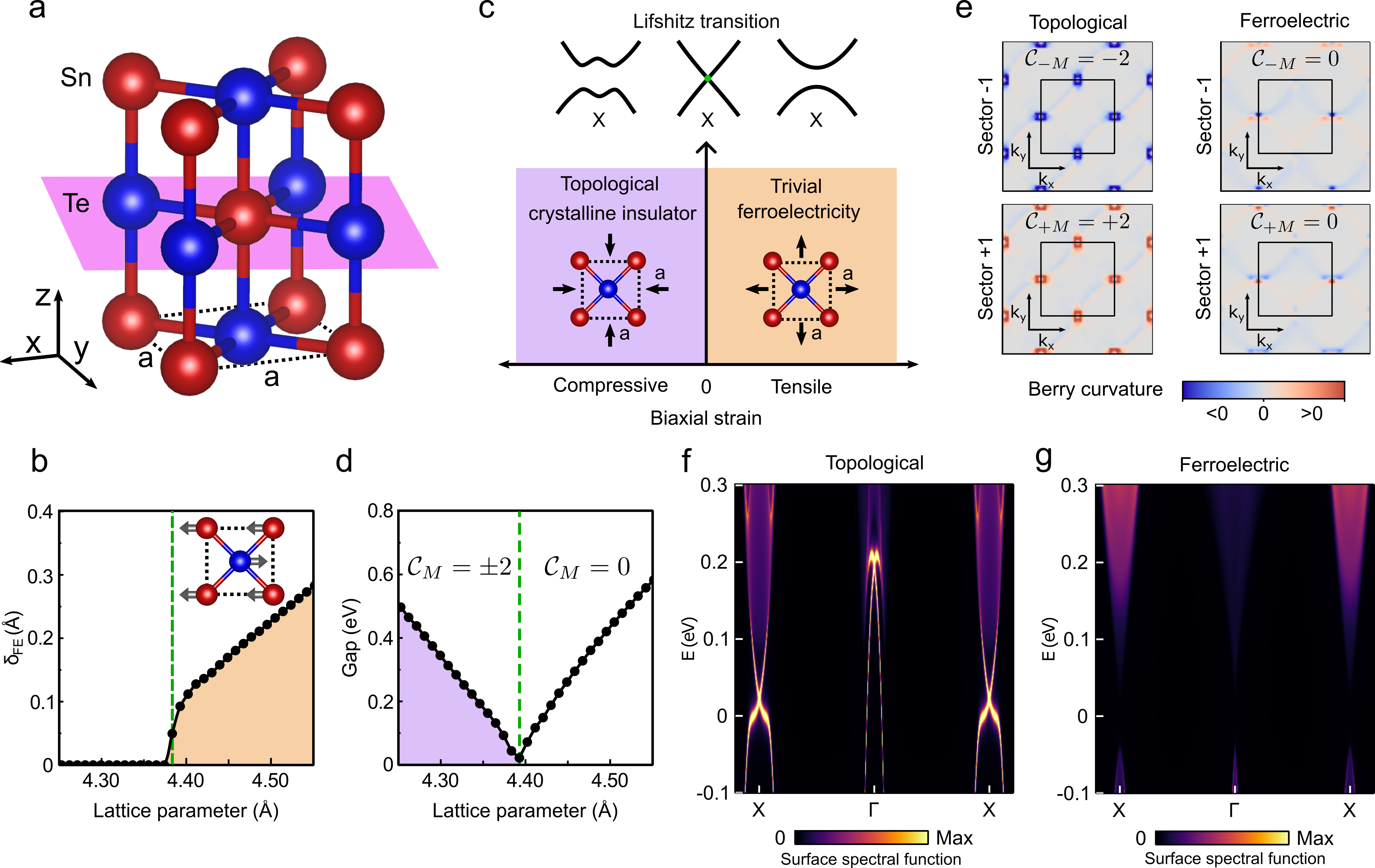}
    \caption{\textbf{Crystal structure and band structure of 3-AL SnTe.} \textbf{a}, Schematic structure of 3-AL SnTe along the (001) orientation, where one mirror plane in the middle of this system from $z\rightarrow-z$ is indicated by a pink shadow. \textbf{b}, Ferroelectric displacement $\delta_{FE}$ between the cation Sn and anion Te evolves with in-plane lattice parameter, schematically represented at the right-top corner. \textbf{c}, Phase diagram of a 3-AL SnTe system as a function of biaxial strain. A Lifshitz transition at the $X$ point takes place when strain goes through from tensile to compressive regime. Correspondingly, the change in Fermi surface topology produces trivial ferroelectric order in tension and a TCI phase in compression. \textbf{d}, Evolution of the band gap at the $X$ point with the lattice constant. The closing of the band gap triggers a topological phase transition that can be defined by time-reversal-mirror Chern number $\mathcal{C}_M$. \textbf{e}, Left(right) plots demonstrate the mirror Berry curvature in reciprocal space for a topological (ferroelectric) phase for the -1 (top) and +1 (bottom) mirror sectors, respectively. \textbf{f} and \textbf{g}, Band structure along $X$-$\Gamma$-$X$ direction for the topological (ferroelectric) phase of a semi-infinite ribbon. Two Dirac points are visible at the $X$ and $\Gamma$ points for the topological phase. }
\label{fig2}
\end{figure}

We now focus on the theoretical analysis of the topological properties in the effective strained 3-AL SnTe system. As shown in Figure \ref{fig2}a, a 3-AL SnTe structural model has a cubic rock-salt structure with a Sn-Sn spacing of 4.5 \AA~ and preserves the reflection symmetry $z\rightarrow-z$ with respect to the atomic layer in the middle plane (pink shadow). This system hosts two equivalent high-symmetry $X$ points in the 2D Brillouin zone. On the other hand, ultrathin SnTe films are known to spontaneously distort along the [111] direction into a rhombohedral structure and form a ferroelectric phase. This in-plane 2D ferroelectricity does not break the mirror symmetry. This allows for competition between ferroelectricity and a TCI phase. \emph{Ab initio} calculations enable us to establish a phase diagram of the 3-AL SnTe system as a function of biaxial strain. Fig. \ref{fig2}b shows ferroelectric displacement $\delta_{FE}$ of this system evolves with the lattice parameter. Above $a\simeq4.38$ \AA, ferroelectric distortion occurs in the $xy$ plane, evidenced by the harmonic phonon spectra (see Extended Data Phonon frequency in SI Figure S5). DFT calculations reveal that the band gap of this system first closes and then reopens at the $X$ point when the lattice constants go through from the tensile to compressive strain, as shown in Fig. \ref{fig2}d. This change in Fermi surface topology corresponds to a Lifshitz transition as a function of in-plane lattice parameter. The gap closing at the $X$ point is the critical point of a phase transition (Fig. \ref{fig2}c), where compressive strain and spin-orbit coupling can induce a band inversion (see Extended Data in SI Figure S3). 

The topological phase is protected by the simultaneous
combination of time-reversal and mirror symmetries, which in the following
we refer to as time-reversal-mirror symmetry. The topological invariant of a strained 3-AL SnTe system can be obtained in the gapped phases by extracting a Wannier Hamiltonian from DFT calculations. The results show that the time-reversal-mirror Chern number vanishes $\mathcal{C}_{\pm M}=0$ in the tensile regime while the time-reversal-mirror Chern number is $\mathcal{C}_{\pm M}=\pm 2$ in the compressive regime (Fig. \ref{fig2}d), signaling the emergence of a TCI phase. This finding is schematically depicted in Fig. \ref{fig2}c.  Fig. \ref{fig2}e displays the computed time-reversal-mirror Berry curvature for each of the symmetry sectors ($\pm 1$) of the operator
$\mathcal{M} = M_{(001)} \Theta$, where $M_{(001)}$ is the $(001)$ mirror and $\Theta$ is the time-reversal symmetry operator. In the topological phase, the mirror Berry curvature is well localized around two $X$ points and possesses an opposite sign for each symmetry sector. In the ferroelectric phase, the mirror Berry curvature is not localized and changes the sign with each symmetry sector, resulting in $\mathcal{C}_{\pm M}=0$. The bulk-boundary correspondence requires that there must be gapless edge states in the biaxially compressive 3-AL system. The time-reversal-mirror Chern number of $\pm 2$ allows two pairs of Dirac cones located around $X$ points in the 2D Brillouin zone, and with further projection into the specific edges, we obtain two pairs of edge state subbands along the $X$-$\Gamma$-$X$ direction (Fig. \ref{fig2}f). Co-propagating edge states carry identical mirror eigenvalues. At the band crossings at $\Gamma$ and $X$ points, edge states with opposite mirror eigenvalues propagate in the opposite direction. The two pairs of edge state subbands exhibit linear dispersion around the crossing points, extending into the bulk band gap. 
In contrast, there are no gapless edge states connecting valence and conduction bands in the ferroelectric phase, as shown in Fig. \ref{fig2}g.

\subsection{\textbf{Experimental evidence of topological edge states}}

\begin{figure}[t!]
    \centering
    \includegraphics[width = 0.9\textwidth]{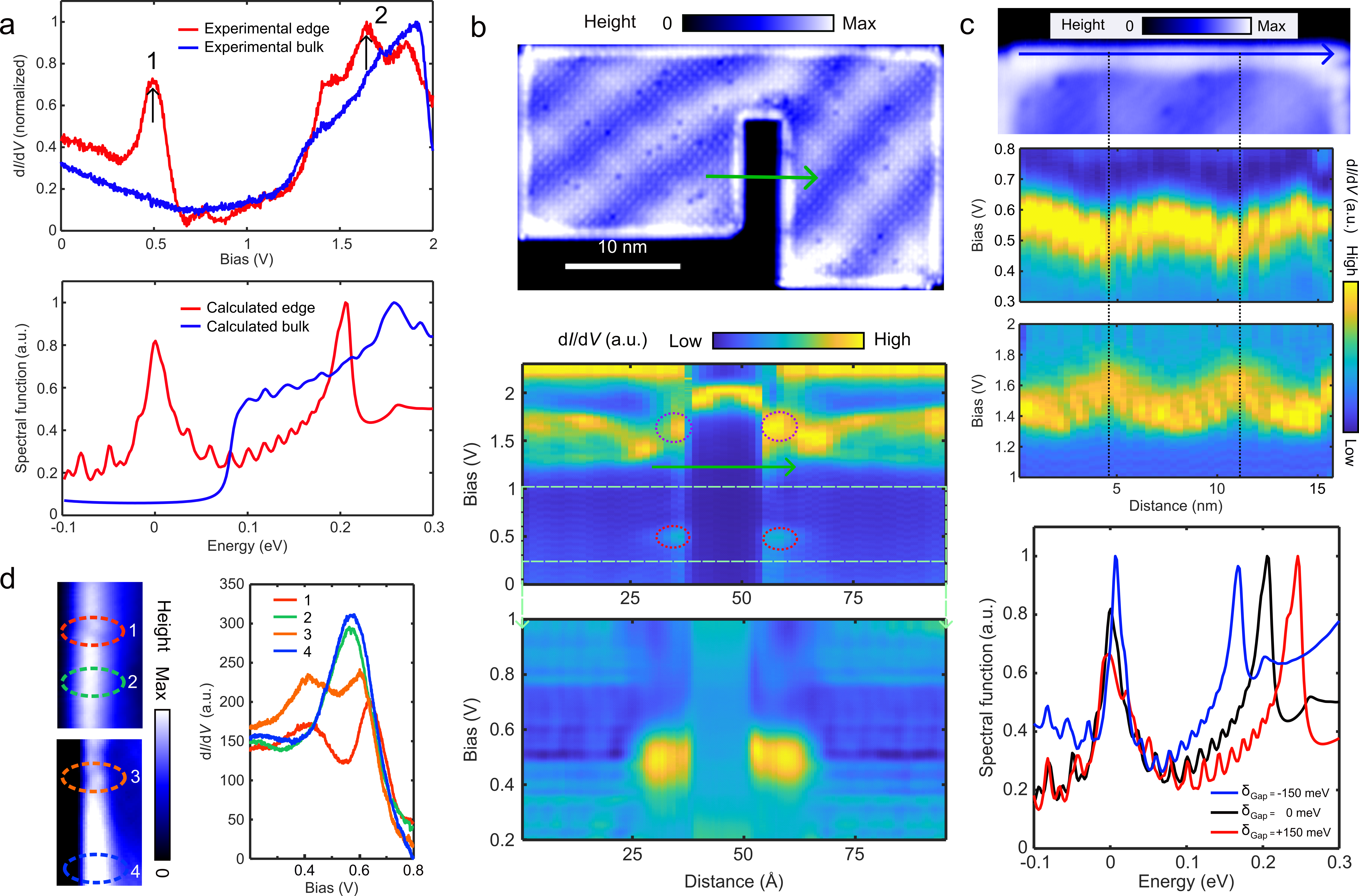}
    \caption{\textbf{The topological edge states and their spatial variation.} \textbf{a}, Tunneling differential spectra taken in the middle and edge of bilayer SnTe on the 2$H$-NbSe$_2$ (top panel) and calculated spectral function of the bulk and edge in the strained 3-AL system (bottom panel). The spectral function is calculated by integrating all the subbands shown in Fig.\ref{fig2}f for a given energy. \textbf{b}, Plots of \dIdV spectra in different energy ranges along the linecut (top panel) are shown in the middle and bottom panel (scan parameters for topography: \textit{V$_b$}=0.8 V, \textit{I}=5 pA). Two pairs of edge state peaks are highlighted by red dashed oval frames. \textbf{c}, Two edge state peaks vary periodically with strained moiré patterns at the edge of bilayer SnTe islands (top and middle panel)(\textit{V$_b$}=0.8 V, \textit{I}=2.2 pA). They have a $\pi$ phase difference in spatial variation. We used a phenomenological method to reproduce this result by calculations (bottom panel). $\delta_{Gap}$ represents the energy gap between two edge state peaks. \textbf{d}, Two typical atomic defects at the edges of bilayer islands are highlighted by red and orange colors and other two ideal sites are also marked by green and blue colors for comparison. Tunneling spectra taken at those positions are shown in the right panel (\textit{V$_b$}=0.8 V, \textit{I}= 5 pA).}
\label{fig3}
\end{figure}

 We present two direct experimental proofs to verify the existence of a 2D TCI in our effective 3-AL system. The first one is that we observe two strong peaks associated to two pairs of edge states extending to the bulk gap along the specific edges of bilayer SnTe islands. As shown in Figure \ref{fig3}a), the \dIdV spectrum taken at the bilayer edge exhibits two pronounced peaks located around 0.5 V (peak 1) and 1.6 V (peak 2). We attribute these peaks to edge states (see SI Figure S10 and S11). These experimental results can be compared with calculations by computing the density of states (DOS) in the bulk and edge of the compressive 3-AL system, where we obtain two sharp peaks with one residing in the bulk gap and the other within the conduction band of bulk states. The low-energy peak at the edge is produced by the substantially enhanced DOS induced by the flatter dispersion of the pair of edge states at the $X$ crossing point (Fig. \ref{fig2}f). Analogously, 
 the high-energy peak at the edge is produced by the enhanced DOS induced by the flatter dispersion of the pair of edge states at the $\Gamma$ crossing point (Fig. \ref{fig2}f). Therefore, each peak corresponds to the crossing of counter-propagating pairs of edge states. The edge state subbands above the peak 1 are fully within the bulk gap (from 0.5 to 0.7 eV). The discrepancy of the energy peak positions between theory and experiment stems from the typical underestimation of the band gap in DFT calculations.

To resolve the spatial modulation of edge states, we first conduct \dIdV spectroscopy across the opposite edges of a bilayer island (top panel in Fig. \ref{fig3}b). At each edge, there are two edge state peaks located at low and high energies, outlined by the dashed oval frames in the middle panel. These peaks are highly localized around the edges. Similar to the observations in Fig. \ref{fig3}a, the peak 1 almost resides the bulk gap (bottom panel) while the peak 2 lies inside the conduction band. It is important to note that the exact energy position of the edge states depends on the strain field, e.g., the strain field can shift the energy of the peak 1 beyond the bulk gap. Series of local DOS (LDOS) oscillations between 0.2 and 1 V as shown in the bottom panel of Fig.~\ref{fig3}b are likely to be related to the LDOS oscillations in the SnTe wetting layer or in the NbSe$_2$ substrate (see further details in the SI Figure S13). We also performed the \dIdV spectroscopy along the blue linecut indicated in Fig. \ref{fig3}c(top panel) and the results show that both the peak 1 and the peak 2 modulate periodically following the moiré pattern at the island edges (two middle panels). Interestingly, the two peaks are anticorrelated, exhibiting a $\pi$ phase difference. The energy shift of the peak 2 is approximately 2 times as large as the peak 1. To rationalize these observations, we developed a phenomenological model. Considering that strain induces changes in the band gap (Fig. \ref{fig2}d), by renormalizing the energy gap as dictated by the strain modulation, we account for the impact of the strain field on the energy position of each pair of edge states. As shown in the bottom panel, our calculations reproduces the experimental results. Moreover, \dIdV maps at different energies (see Extended Data in SI Figure S17) reveal that the two pairs of edge states span the bulk energy gap along the perimeters of bilayer islands at $0.5$ and $0.6$ V. They only emerge along the sharp and straight edges with specific atomic terminations of either Sn or Te atoms, a characteristic observed in both regular and irregular bilayer SnTe structures (see Extended Data in SI Figures S19 and S20). 

Since the edge states in our system are protected by time-reversal-mirror symmetry, the second piece of evidence to confirm a 2D TCI is to observe a band gap in the edge states when time-reversal-mirror symmetry is broken. Time-reversal-mirror symmetry can be disrupted spontaneously or through external perturbations such as perpendicular electric field or an in-plane magnetic field\cite{Liu2014}. In contrast with time-reversal symmetry, disorder can always break time-reversal-mirror symmetry. In our effective 3-AL system, intrinsic atomic defects at the edges locally and randomly break the $z\rightarrow-z$ mirror symmetry about the middle plane. As a result, the edge states open up a band gap, as shown in Fig. \ref{fig3}d. The depth of the opened band gap depends on the extent to which mirror symmetry is broken at the edges. Remarkably, we find that the edge states are substantially resilient to the effects of the substrate, the moiré potential, and even a sizable electric field (see Extended Data in SI Figure S15 and S16).

\subsection{\textbf{Coupling of topological edge states}}

\begin{figure}[t!]
    \centering
    \includegraphics[width = 0.9\textwidth]{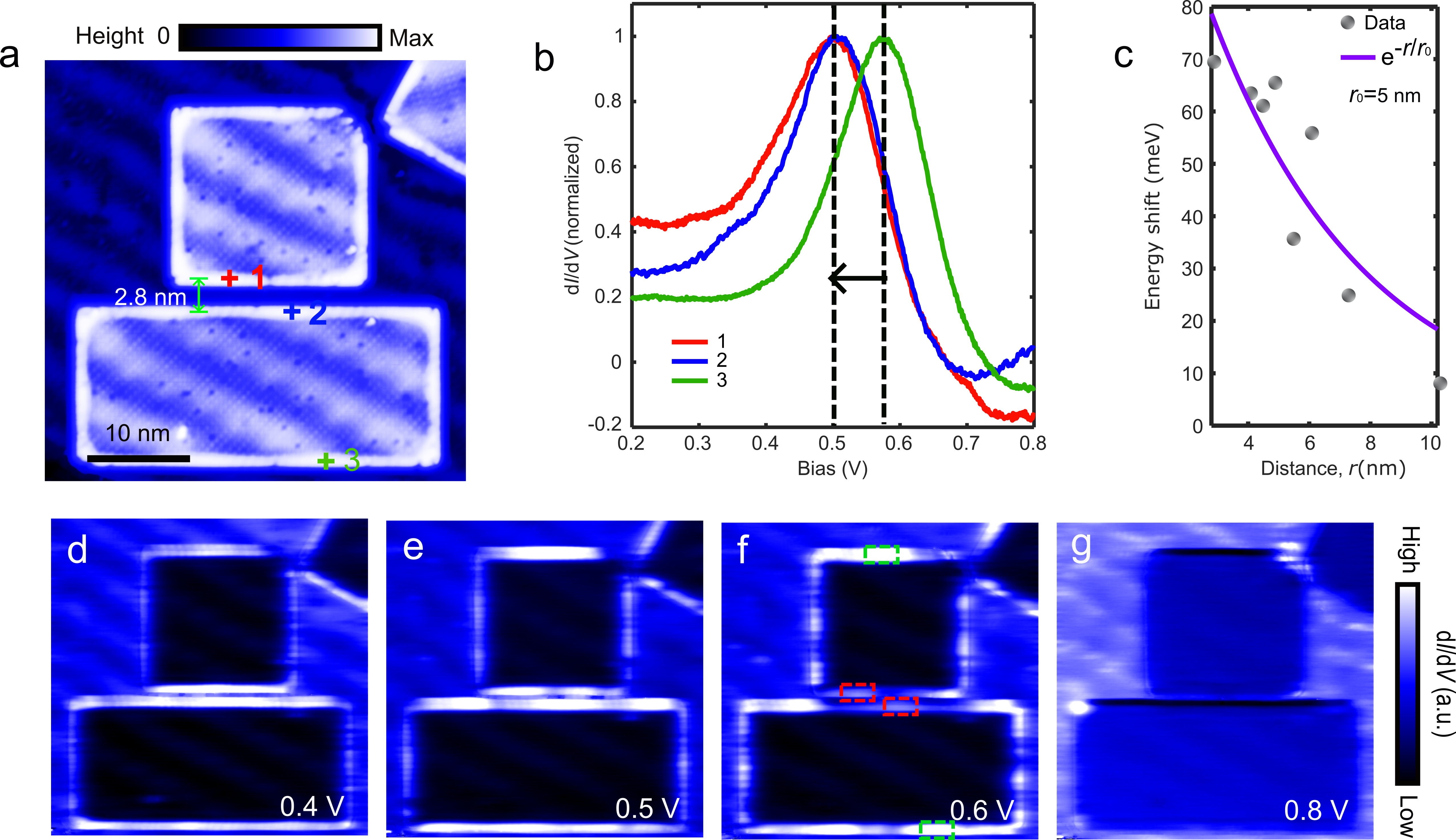}
    \caption{\textbf{Coupling of topological edge channels.} \textbf{a}, STM topographic image of two adjacent bilayer SnTe islands (\textit{V$_b$}=0.5 V, \textit{I}=60 pA). The distance of their neighboring edges is 2.8 nm. \textbf{b}, Tunneling spectra taken at the equivalent points 1, 2 and 3 from three edges in \textbf{a} by locating the local maximum position of the edge state peaks. The black arrow indicates the shift direction of neighboring edge states. \textbf{c}, The blue curve fits the exponential decay of the energy shift of edge states with the increasing distance between two adjacent edges. The decay length is 5 nm. \textbf{d}-\textbf{g}, \dIdV maps over the bilayer island from \textbf{a} at different bias voltages. The red and green dashed rectangular frames outline the edge states with and without neighboring edges. The tunneling junction setup is \textit{I$_{set}$}=60 pA.}  
\label{fig4}
\end{figure}

The interaction of edge states in 3D TCIs has been studied\cite{doi:10.1126/science.aah6233, PhysRevLett.126.236402}. We probe the possibility of observing coupling between the edge states in our 2D TCI by examining two closely adjacent bilayer islands, as illustrated in Figure \ref{fig4}a. High-resolution topography indicates no significant structural distortions or atomic defects that could disrupt mirror symmetry, so a gap opening in the edge states is not expected. To eliminate the influence of strain, which can shift edge state energy, the tunneling spectra were taken at equivalent points on the same stripe pattern, ensuring a consistent strain field. Our \dIdV spectra reveal that peak 1 shifts to lower energy compared to isolated edges as indicated by the black arrow (Fig. \ref{fig4}b) and its intensity weakens significantly around 0.6 V. The energy shift due to the proximity of adjacent edges exhibits an exponential decay with increasing distance between the two edges, as seen in Fig. \ref{fig4}c, with a decay length of approximately 5 nm. This effect is most evident in the \dIdV mapping over the bilayer edges at different energies (Figs. \ref{fig4}d-g), as outlined by the green and red rectangular frames. 

We propose two possible origins for the energy shift with both expected to coexist in our system. First, we experimentally observe that the edge states shift towards lower energy compared to isolated edges, which can be explained by electrostatic coupling. This type of coupling can alter the energy position of edge states and is sensitive to both the distance between the edges and the strength of the electric field or charge between adjacent edges. Second, there is a possibility of substrate-mediated tunneling between adjacent edges, allowing for interaction through proximity. To assess the magnitude of this effect, we developed a model where the electronic modes in one bilayer island can tunnel into another island through an effective barrier potential $V$ (details in SI Figure S22). This tunneling coupling leads to a shift of the crossing Dirac point. Comparison with the experimental results is possible by relating $V$ to the distance $d$ between neighboring edges through $\lambda d =\ln\left(\frac{V}{t_0}\right)$, where $t_0$ is the hopping energy of SnTe orbitals (on the order of eV) and $\lambda$ is a finite phenomenological parameter that can be adjusted based on experimental data. This model results in an exponential decay of the energy shift as $\lambda d $ increases, consistent with our experimental results. In summary, it is likely that a combination of electrostatic interactions and tunneling coupling is responsible for the energy shift of edge modes for closely spaced SnTe edges.

\section{Conclusion}

In conclusion, we experimentally realized a 2D TCI by artificially constructing a heterostructure of bilayer SnTe on the 2$H$-NbSe$_2$ substrate. First, we show that this experimental system behaves as an effective strained 3-AL SnTe structure in a compressive regime.\emph{Ab initio} calculations unveil that compressive strain drives this system through a phase transition from a trivial ferroelectric to a crystalline topological insulator that hosts two pairs of edge states protected by time-reversal-mirror symmetry. Second, we employ STM and STS to observe two pronounced peaks associated with the two pairs of edge states within fully gapped bulk states and visualize their propagation along the edges of bilayer islands. Intrinsic atomic defects locally break the time-reversal-mirror symmetry and thus open up a band gap in the edge states. Finally, we show that two closely adjacent topological edge modes can be shifted by a combination of electrostatic interactions and tunneling coupling. Our results establish strain as a powerful control parameter and demonstrate that it can be used to drive few-layer SnTe to a 2D crystalline topological insulator. This crystalline topological phase provides a starting point to potentially engineer Chern insulators by including ferromagnetically ordered dopants\cite{Yu2010,PhysRevLett.112.046801,Bernevig2022} or to use this as a building block in van der Waals heterostructures, for example, targeting topological mirror superconductivity\cite{PhysRevLett.111.056403,PhysRevLett.100.096407} by proximity effect.

\section{Experimental Methods}

Ultrathin SnTe films were grown on the freshly cleaved bulk crystal 2$H$-NbSe$_2$ by molecular beam epitaxy under ultrahigh vacuum conditions (UHV, base pressure $\approx 2\times 10^{-10}$ mbar). Before the growth, compound material SnTe was fully degassed and 2$H$-NbSe$_2$ substrate was also out-gassed at $\sim100^\circ$C. During the growth, the SnTe powder of 99.999 \% purity loaded into a Knudsen cell was evaporated onto the substrate held at room temperature for 3 minutes. After the growth, the sample was annealed at $\sim50^\circ$C. Finally, it was inserted into low-temperature STM head housed in the same UHV system. All the subsequent STM and STS measurements were performed at T = 4.7 K. The etched tungsten tip was used for all the STM scan and electrical measurement. STM images were scanned in the constant-current mode and all the \dIdV spectra were recorded by a standard lock-in techniques while sweeping the sample bias in an open feedback loop configuration, with a peak-to-peak bias modulation of 5$\sim$20 mV at a frequency of 746 Hz.

The codes used to extract strain map of bilayer SnTe are based on the available library (https://github.com/LowTemperaturesUAM/Strain-calculation-from-2D-images). 

\section{Theoretical Methods}

DFT calculations were realized with the standard Perdew-Burke-Ernzerhof (PBE) functional \cite{PhysRevLett.77.3865} as implemented as in the QUANTUM-ESPRESSO package \cite{Giannozzi2009}. Ultra-soft pseudopotentials from the pslibrary (https://dalcorso.github.io/pslibrary/) version 1.0.0 were used to describe the interactions between electrons and ions \cite{DALCORSO2014337}, while the wave functions were expanded using a plane-wave basis set with a kinetic energy cutoff of 50 Ry. The integration over the Brillouin zone was performed using a 8$\times$8$\times$1 grid. After structural relaxation with the force threshold of 0.01 eV/Å, all electronic properties were obtained considering spin-orbit coupling with the full-relativistic versions of the same pseudopotentials used previously and Brillouin zone sampling of 10$\times$10$\times$1. The Wannier90 package \cite{Pizzi2020} was employed to obtain tight-binding Hamiltonians in the basis of maximally localized Wannier functions, including $s$ and $p$ orbitals for both Sn and Te atoms. We fitted the DFT band structures with the tight-binding Hamiltonians considering an energy window of more than 20 eV and a frozen window covering 2 eV. 

The same code and pseudopotentials were used to relax and to obtain electronic properties of SnTe/NbSe$_2$ heterostructures. The Brillouin zone was integrated using a smaller 2$\times$2$\times$1 grid, and the vdw-df-ob86 vdW functional was used instead of standard PBE in order to account for weak dispersion interactions that might occur between the layers \cite{PhysRevB.83.195131}. All calculations in this part were performed without taking spin-orbit coupling interactions into account.

\section*{Acknowledgements}
This research made use of the Aalto Nanomicroscopy
Center (Aalto NMC) facilities and was supported by
the European Research Council (ERC-2021-StG
No. 101039500 “Tailoring Quantum Matter on the
Flatland” and ERC-2017-AdG No. 788185 “Artificial
Designer Materials”) and Academy of Finland (Academy
Research Fellow No. 331342, No 358088, No. 336243, No. 338478,
No. 346654, Academy postdoctoral fellow No. 349696, and the Finnish Quantum Flagship, No. 358877). We acknowledge the computational resources provided by the Aalto
Science-IT project.
\section*{Author contributions}
L.J., M.A., S.K., J.L.L., and P.L. conceived the experiment, and L.J., M.A., carried out the sample growth and the low-temperature STM experiments. L.J., M.A., A.O.F., and O.J.S. analysed the STM data. A.O.F.,  O.J.S., and J.L.L. developed the theoretical model. L.J., A.O.F., O.J.S., S.K., J.L.L., and P.L. wrote the manuscript with input from all co-authors.

%\PL{Who wants to write something real here?}

\section*{Competing interests}
The authors declare no competing interests.

\bibliography{Ref}

\end{document}

\title{Supplementary information for strain-induced two-dimensional topological crystalline insulator 
\date{\today}
\author{Liwei Jing}
\affiliation{Department of Physics, Nanoscience Center, 
University of Jyväskyl\"a, FI-40014 University of Jyväskyl\"a, Finland}

\author{Mohammad Amini}
\thanks{These authors contributed equally to this work.}
\affiliation{Department of Applied Physics, Aalto University, FI-00076 Aalto, Finland}

\author{Adolfo O. Fumega}
\thanks{These authors contributed equally to this work.}
\affiliation{Department of Applied Physics, Aalto University, FI-00076 Aalto, Finland}

\author{Orlando J. Silveira}
\thanks{These authors contributed equally to this work.}
\affiliation{Department of Physics, Nanoscience Center, 
University of Jyväskyl\"a, FI-40014 University of Jyväskyl\"a, Finland}
\affiliation{Department of Applied Physics, Aalto University, FI-00076 Aalto, Finland}

\author{Jose L. Lado}
\email{e-mail: kezilebieke.a.shawulienu@jyu.fi; jose.lado@aalto.fi; peter.liljeroth@aalto.fi}
\affiliation{Department of Applied Physics, Aalto University, FI-00076 Aalto, Finland}

\author{Peter Liljeroth}
\email{e-mail: kezilebieke.a.shawulienu@jyu.fi; jose.lado@aalto.fi; peter.liljeroth@aalto.fi}
\affiliation{Department of Applied Physics, Aalto University, FI-00076 Aalto, Finland}

\author{Shawulienu Kezilebieke}
\email{e-mail: kezilebieke.a.shawulienu@jyu.fi; jose.lado@aalto.fi; peter.liljeroth@aalto.fi}
\affiliation{Department of Physics, Nanoscience Center, 
University of Jyväskyl\"a, FI-40014 University of Jyväskyl\"a, Finland}
\affiliation{Department of Chemistry and Nanoscience Center, 
University of Jyväskyl\"a, FI-40014 University of Jyväskyl\"a, Finland}

\maketitle

\section{DFT calculations - a few atomic layers of freestanding SnTe}

\subsection{Band-structure calculations}

Figure \ref{symgeo}\textbf{a} reveals the mirror, or symmorphic, symmetry of a few-layer SnTe when the number of atomic layers (ALs) is odd, allowing the definition of the mirror Chern number. Adding another AL makes the SnTe non-symmorphic symmetric (Fig.~\ref{symgeo}\textbf{b}), which does not allow to define a mirror Chern number.

\begin{figure*}[h!]
    \centering
    \includegraphics[width = 0.9\textwidth]{SI Figures/symgeo.png}
    \caption{Schematic view of 3 ALs \textbf{(a)} and 4 ALs \textbf{(b)} of SnTe stacked along the (001) direction. Red balls: Sn atoms and blue balls: Te atoms.}
\label{symgeo}
\end{figure*}
\newpage

Figures \ref{nostrain_bands}\textbf{(a)}-\textbf{(e)} reveal that the non-symmorphic symmetry gives an extra degeneracy to the band structure at the X point, significantly increasing the band gap in the few-layer limit. Figs. \ref{nostrain_bands}\textbf{(f)}-\textbf{(j)} show that the degeneracy from non-symmorphism is intact against spin-orbit coupling interaction.

\begin{figure*}[h!]
    \centering
    \includegraphics[width = 0.9\textwidth]{SI Figures/nostrain_bands.png}
    \caption{\textbf{(a)}-\textbf{(e)}, Band structures of SnTe with 1, 2, 3, 4 and 5 ALs obtained with their relaxed lattice parameters keeping the tetragonal symmetry. \textbf{(f)}-\textbf{(j)}, Band structures after considering spin-orbit coupling interaction, where green and orange colors indicate the projections onto the p orbitals of Sn and Te atoms, respectively.}
\label{nostrain_bands}
\end{figure*}

Perfectly planar 1-AL SnTe has a nontrivial band inversion with non-zero mirror Chern number, as seen in Figs.~\ref{nostrain_bands} \textbf{(a)} and \textbf{(f)}. Adding more atomic layers introduces a hybridization between the top and bottom atomic layers, destroying the band inversion even in the symmorphic cases.

\newpage
Figure \ref{3layersbands} shows that reducing the lattice parameter induces band inversion in the case of 3 ALs, for example, having inverted contributions from Sn and Te orbitals around the X point.

\begin{figure*}[h!]
    \centering
    \includegraphics[width = 0.95\textwidth]{SI Figures/3layerbands.png}
    \caption{Band structures of the 3 layers with different biaxial strain and considering spin-orbit coupling. Green and orange colors indicate the projections onto the p orbitals of Sn and Te atoms, respectively.}
\label{3layersbands}
\end{figure*}

The kind of band inversion shown in Fig.~\ref{3layersbands} is a sign of nontrivial topology, and we calculated the mirror Chern number and edge states based on the Wannier Hamiltonian of the compressed 3-AL SnTe with lattice parameter of 4.33 \r{A}. The band structures of the 2D SnTe stripes along the (100) direction with different widths are shown in Figure \ref{gap_evolv}. In Figures~\ref{3layersbands} and Fig.~\ref{gap_evolv} the 3-AL SnTe is metallic due to a hole pocket around the $X$ point, preventing a definition of a topological invariant. However, the band structure of the stripes clearly show that edge bands emerge within the region where the band inversion happens. DFT is well known to underestimate gaps and our experiments show a bulk excitation gap in this system.
To account for this effect, 
we calculated the band structures of the 2D system using a Wannier Hamiltonian computed with a band gap in Bloch Hamiltonian
that is corrected with a scissors operator before the Wannierization. 
The scissors operator does not allow for any band crossing, thus not inducing any topological phase transition while lifting the hole pocket around $X$ (see Fig.~\ref{gap_evolv}). The resulting edge states from the shifted Wannier Hamiltonian reside within the current global band gap, and stem
from the mirror Chern number computed from the Wannier Hamiltonian $C = \pm 2$.\\

\begin{figure*}[h!]
    \centering
    \includegraphics[width = 0.95\textwidth]{SI Figures/gap_evolv.png}
    \caption{\textbf{a} and \textbf{b}, Wannier band structures of the compressed 3-AL SnTe with a = 4.33 \r{A}. \textbf{c} and \textbf{d}, Wannier bands of the same system obtained after shifting the conduction and valence bands by $\pm$ 0.2 eV. The green arrows in \textbf{a}$_{\text{vii}}$ and \textbf{b}$_{\text{vii}}$ indicate the hole pocket around $X$, while the black arrows indicate the edge states. \textbf{d}$_{\text{vii}}$ (\textbf{a}$_{\text{i}}-$\textbf{a}$_{\text{vii}}$) and (\textbf{b}$_{\text{i}}-$\textbf{b}$_{\text{vii}}$), Band structure of (100) ribbons with projections onto two different types of edge terminations shown in blue and red. (\textbf{c}$_{\text{i}}-$\textbf{c}$_{\text{vii}}$) and (\textbf{d}$_{\text{i}}-$\textbf{d}$_{\text{vii}}$), Band structure of the same systems but with the shifted bands. The black arrows in \textbf{c}$_{\text{vii}}$ and \textbf{d}$_{\text{vii}}$ indicate the edge states, now isolated within the bulk band gap. \textbf{(e)}, Example of a ribbon geometry with the periodicity along the (100) direction and with a vacuum layer along the (010) direction. The edge configurations are defined by the terminated atoms at the topmost edge, which are either Sn (red) or Te (blue) atoms.}
\label{gap_evolv}
\end{figure*}

To further confirm the existence of ferroelectricity in the 3-AL system, we computed the harmonic phonon spectrum as a function of lattice constant (Figure \ref{Phonon frequency}). As the phonon mode is related to the ferroelectric displacement at the $\Gamma$ point, negative (positive) frequencies correspond to an imaginary unstable (stable) structure. This evolution suggests that the ferroelectric order only exists in biaxially tensile 3-AL system.

\begin{figure*}[h!]
    \centering
    \includegraphics[width = 0.5\textwidth]{SI Figures/Phonon frequency.png}
    \caption{Evolution of ferroelectric phonon modes with in-plane lattice parameter. $\omega$\textsubscript{FE} represents the phonon frequency of ferroelectricity.} 
\label{Phonon frequency}
\end{figure*}
\newpage

\subsection{ Computation of the mirror Chern number}

The topological invariant of our topological crystalline insulator (TCI) is given by the time-reversal-mirror Chern number, that can be computed as follows.
The Bloch Hamiltonian has a set of associated set of eigenstates given by

\begin{equation}
\mathcal{H}_{\mathbf k} | \Psi_\alpha (\mathbf k) \rangle = 
\epsilon_{\mathbf k} | \Psi_\alpha (\mathbf k) \rangle
\end{equation}

where $| \Psi_\alpha (\mathbf k) \rangle$ are the Bloch wavefunctions and $\epsilon_{\mathbf k}$ are the Bloch eigenvalues.
The Hamiltonian $\mathcal{H} (\mathbf k)$ commutes with the time-reversal mirror operator $\mathcal{M} = M_z \Theta$ as 

\begin{equation}
[\mathcal{H}_{\mathbf k},\mathcal{M}] = \mathcal{H}_{\mathbf k}\mathcal{M} - \mathcal{M} \mathcal{H}_{\mathbf k} =0
\end{equation}

which allows to define a basis where the $| \Psi_\alpha (\mathbf k) \rangle$ are simultaneously eigenstates of the Hamiltonian
and the time-reversal-mirror operator. In particular, in such a basis we have

\begin{equation}
\mathcal{M}| \Psi_\alpha (\mathbf k) \rangle = 
\chi | \Psi_\alpha (\mathbf k) \rangle
\end{equation}

where $\chi = 1$ for $\alpha \in M^+$ and $\chi = - 1$ for $\alpha \in M^-$. Having classified the Bloch states in two sets with different
symmetry eigenvalues, the time-reversal-mirror Chern number can be computed as

\begin{equation}
\mathcal{C}_M =
\frac{1}{2\pi} \int \Omega^+ d^2 \mathbf k - 
\frac{1}{2\pi} \int \Omega^- d^2 \mathbf k 
\end{equation}

where the time-reversal-mirror Berry curvature is computed with a Wilson loop in each time-reversal-mirror sector

\begin{equation}
\Omega^{\pm} (\mathbf k) = \lim_{\Delta k \rightarrow 0}\frac{1}{(\Delta k)^2} \text{arg} \left (\text{det} [U^{\pm}_{1,2} (\mathbf k)  U^{\pm}_{2,3} (\mathbf k) U^{\pm}_{3,4} (\mathbf k) U^{\pm}_{4,1} (\mathbf k)] \right )
\end{equation}

where 
$U^{\pm}_{1,2} (\mathbf k) = \langle \Psi_\alpha (k_x,k_y) |\Psi_\beta (k_x +\Delta k,k_y) \rangle $,
$U^{\pm}_{2,3} (\mathbf k) = \langle \Psi_\alpha (k_x + \Delta k,k_y) |\Psi_\beta (k_x +\Delta k,k_y+ \Delta k) \rangle $,
$U^{\pm}_{3,4} (\mathbf k) = \langle \Psi_\alpha (k_x +\Delta k,k_y+ \Delta k) |\Psi_\beta (k_x,k_y+ \Delta k) \rangle $,
$U^{\pm}_{4,1} (\mathbf k) = \langle \Psi_\alpha (k_x,k_y+ \Delta k) |\Psi_\beta (k_x,k_y) \rangle $, 
where $\alpha,\beta \in M^+$ for $U^+$ and $\alpha,\beta \in M^-$ for $U^-$.
The previous methodology allows us to compute the time-reversal-mirror Berry curvature and
time-reversal-mirror Chern number directly from the first principles electronic structure through
a Wannierization procedure.

\newpage
\section{Growth of bilayer SnTe}

\begin{figure*}[h!]
    \centering
    \includegraphics[width = 0.9\textwidth]{SI Figures/STM topography.png}
    \caption{\textbf{SnTe structure on 2$H$-NbSe$_2$ and graphite.} \textbf{a}, STM image of bilayer SnTe on the 2$H$-NbSe$_2$ substrate (tunneling parameters: \textit{V$_b$}=1V, \textit{I}=3.3 pA). Monolayer islands exhibit different dislocation orientations and the bilayer follows the underlying monolayer moiré pattern. \textbf{b}, Linecut from the substrate to the top layer in SnTe islands (\textit{V$_b$}=1 V, \textit{I}=3.4 pA). The corresponding height profile in \textbf{c} suggests that our system is composed of bilayer SnTe since each monolayer in the rock-salt structure is 0.632 nm thick. \textbf{d}, Atomic-scale STM image of monolayer SnTe shows a buckled and disordered structure, which is reflected by two sets of different Bragg peaks outlined by green and orange dashed circles via fast Fourier transform in the inset (\textit{V$_b$}=0.5 V, \textit{I}=50 pA). This is a consequence of strong interaction and large lattice mismatch between the monolayer SnTe and the substrate. \textbf{e}, High-resolution STM image of bilayer SnTe (\textit{V$_b$}=0.5 V, \textit{I}=100 pA). The sublattice Sn atoms aligned along two orthogonal directions are normal to the edge. \textbf{f}, Large-scale STM image (in the top: \textit{V$_b$}=1.6 V, \textit{I}=12.1 pA) and atomically resolved topographic image (in the bottom: \textit{V$_b$}=0.077 V, \textit{I}=256 pA) show that Sn atoms are aligned along [110] and [1$\bar{1}$0] directions for the monolayer SnTe on highly oriented pyrolytic graphite (HOPG), which favors the in-plane ferroelectricity.}
    \label{STM topgraphy}
\end{figure*}

\section{\textbf{Spatially varying strain in bilayer SnTe}}

\begin{figure*}[h!]
    \centering
    \includegraphics[width = 0.9\textwidth]{SI Figures/strain map.png}
    \caption{\textbf{a}, STM topography cropped from the Figure 1\textbf{c} in the main text for analysis to avoid the edge effects (the image size: 23 nm by 23 nm), and the corresponding biaxial strain map is shown in \textbf{b}. \textbf{c}, High-resolution STM topographic image by cropping the central part of the bilayer island shown in Figure~\ref{STM topgraphy}\textbf{e}(12.4 nm by 12.4 nm). \textbf{d}, Corresponding biaxial strain map. }   
\label{strain map}
\end{figure*}

To understand the impact of strain on the electronic structure of bilayer SnTe, we apply the well-established Lawler-Fujita algorithm\cite{Lawler2010} to the atomically resolved topography shown in Figure~\ref{strain map}a and determine strain map at the nanoscale. Since some effects contribute to the $\textbf{u}(\textbf{\textit{r}})$, such as thermal drift, piezo relaxation/hysteresis and nonlinearity, we extract the intrinsic strain field by subtracting a 3$^{rd}$ polynomial fit in $x$ and $y$ components of $\textbf{u}(\textbf{\textit{r}})$, which sufficiently describes those effects varying slowly in space. By grabbing the reciprocal unit-cell Bragg peaks $\textbf{\textit{Q}}_a$ and $\textbf{\textit{Q}}_b$ by Fourier transform, we can determine the two orthogonal wave vectors reflecting the atomic corrugations in tetragonal SnTe structure. Hence, the ideal topographic image is expressed by 

\begin{equation}   
\textit{T}(\textbf{\textit{r}})=T_0[\cos(\textbf{\textit{Q}}_a\cdot\textbf{\textit{r}})+\cos(\textbf{\textit{Q}}_b\cdot\textbf{\textit{r}})].
\end{equation}
The coordinates in a square array of pixels are labeled \textbf{\textit{r}}=$(x,y)$. The real topography with lattice distortion is schematically written by the varying field as
\begin{equation}
\textit{T}(\textbf{\textit{r}})=T_0[\cos(\textbf{\textit{Q}}_a\cdot(\textbf{\textit{r}}+\textbf{u}(\textbf{\textit{r}})))+\cos(\textbf{\textit{Q}}_b\cdot(\textbf{\textit{r}}+\textbf{u}(\textbf{\textit{r}})))].
\end{equation}
On the other side, the lattice distortion can be equivalently related to the local phase of the atomic cosine components in the following,

\begin{align}
\label{eqn:eqlabel1} 
\begin{split}
\textbf{\textit{Q}}_a\cdot\textbf{\textit{r}}+\theta_a(\textbf{\textit{r}})&=\textbf{\textit{Q}}_a\cdot\textbf{\textit{r}}^{\prime}
\\
\textbf{\textit{Q}}_b\cdot\textbf{\textit{r}}+\theta_b(\textbf{\textit{r}})&=\textbf{\textit{Q}}_b\cdot\textbf{\textit{r}}^{\prime}.
\end{split}
\end{align}

Here $\textbf{\textit{r}}^{'}$ is the point of equal phase on the perfect lattice periodic image. 
Then the strain field is expressed by

\begin{equation}
\begin{pmatrix}
    \theta_a(\textbf{\textit{r}})\\
    \theta_b(\textbf{\textit{r}})
\end{pmatrix}=-
\begin{pmatrix}
    \textbf{\textit{Q}}_a\\
    \textbf{\textit{Q}}_b
\end{pmatrix}\cdot
(\textbf{\textit{r}}-\textbf{\textit{r}}^{\prime})
\end{equation}
Since $\textbf{\textit{Q}}_a$ and $\textbf{\textit{Q}}_b$ are orthogonal and invertible, we obtain the displacement field:
\begin{equation}
    \textbf{\textit{u}}(\textbf{\textit{r}})=\textbf{\textit{r}}-\textbf{\textit{r}}^{\prime}=-\textbf{\textit{Q}}^{-1}
    \begin{pmatrix}
    \theta_a(\textbf{\textit{r}})\\
    \theta_b(\textbf{\textit{r}})
    \end{pmatrix}
\end{equation}
We now extract the local phase information $\theta_i$(\textbf{\textit{r}}) by applying a spatial lock-in technique to the topography. First, we introduce the coarsening length scale $\frac{1}{\lambda}$, over which $\textbf{u}(\textbf{\textit{r}})$ changes very slowly. Then we choose very small $\lambda$ due to quite sharp lattice peaks. With a low-pass filter function, we expand the topography as 

\begin{align}
\label{eqn:eqlabel2} 
\begin{split}
 \textit{T}_x(\textbf{\textit{r}})&=\sum_{\textbf{\textit{r}}^{\prime}}\textit{T}(\textbf{\textit{r}}^{\prime})e^{-i\textbf{\textit{Q}}_x\cdot\textbf{\textit{r}}^{\prime}}\frac{\lambda^2}{2\pi}e^\frac{-\lambda ^2 (\left|\textbf{\textit{r}}-\textbf{\textit{r}}^{\prime}\right|)^2}{2}\\
\textit{T}_y(\textbf{\textit{r}})&=\sum_{\textbf{\textit{r}}^{\prime}}\textit{T}(\textbf{\textit{r}}^{\prime})e^{-i\textbf{\textit{Q}}_y\cdot\textbf{\textit{r}}^{\prime}}\frac{\lambda^2}{2\pi}e^\frac{-\lambda ^2 (\left|\textbf{\textit{r}}-\textbf{\textit{r}}^{\prime}\right|)^2}{2}.
\end{split}
\end{align}

Then $\theta_a$ and $\theta_b$ are given by
\begin{equation}
    \theta_i(\textbf{\textit{r}})=\arctan\left(\frac{\Im{\textit{T}_i(\textbf{\textit{r}})}}{\Re{\textit{T}_i(\textbf{\textit{r}})}}\right).
\end{equation}

As the $\arctan$ function can have multiple values, we take a derivative of the image to locate the jumps and add one 2$\pi$ to make $\textbf{u}(\textbf{\textit{r}})$ a single valued quantity. Finally, with the above information and the elasticity theory\cite{10.1063/1.3057037}, we obtain the displacement field $\textbf{u}(\textbf{\textit{r}})$ and the distribution of biaxial strain $s$ given by 
\begin{equation}
\textit{s}=-\frac{1}{2}(u_{xx}+u_{yy}),
\end{equation}
where the partial derivatives \textit{u}$_{xx}$=$\partial_xu_x$ and \textit{u}$_{yy}$=$\partial_yu_y$ are effective strain components. The negative values in the strain map shown in Fig.~\ref{strain map}b and d represent the compressive strain while the positive values denote the tensile strain.

\section{DFT calculations - heterostructures}

We performed DFT calculations of $x$-SnTe/NbSe$_2$ heterostructures, where $x=1$ to $4$ indicates the layer number of SnTe including the interaction with one monolayer NbSe$_2$. Figure~\ref{DFT-hetero}\textbf{a} shows the unit cell of one monolayer NbSe$_2$  and the cell used to create the heterostructures that imposes the minimum strain to both SnTe and NbSe$_2$. Fig.~\ref{DFT-hetero}\textbf{b} shows a SnTe cell as an example, and Figs. \ref{DFT-hetero}\textbf{c} and\textbf{ d} show the top and side view of the relaxed 4-SnTe/NbSe$_2$ heterostructures (keeping a square cell), where the compressive strain in SnTe is given by $\epsilon$ and tensile strain is given by $\eta$. Figs. \ref{DFT-hetero}\textbf{e}-\textbf{h} show another example of the heterostructure that produces a sizeable compressive strain to the SnTe while keeping the same NbSe$_2$ layer as before. When $x$=4, for example, the compression of $\epsilon_y = 9.2 \% $ and $\epsilon_x = 8.2 \% $ occurs in the 4-AL SnTe, which leads to a large distortion in the bottom 1-AL SnTe in contact with the NbSe$_2$ monolayer, as illustrated in Fig.~\ref{DFT-hetero}\textbf{h}.

\begin{figure*}[h!]
    \centering
    \includegraphics[width = 0.9\textwidth]{SI Figures/DFT-hetero.png}
  \caption{\textbf{a}-\textbf{h}, Heterostructure consisting of the blue cell of NbSe$_2$ \textbf{a} and the black cell of a 4-AL SnTe \textbf{b}. The heterostructure was fully relaxed (both the length and angles of the cell), resulting in the structure shown in \textbf{c} and \textbf{d}. \textbf{e}-\textbf{f}, Same procedures but with a slightly larger cell of 4-AL SnTe rotated by 45$\degree$ in \textbf{f}. Only the atomic positions were relaxed while the cell was kept fixed, resulting in the structure shown in \textbf{g} and \textbf{h}. The side views in \textbf{d} and \textbf{h} were positioned in a way to show clearly the displacement of each atom along the $z$ direction. }
\label{DFT-hetero}
\end{figure*}

In both scenarios shown in Figs. \ref{DFT-hetero}d and h, the tensile strain on the NbSe$_2$ layer was kept minimum ($\eta = 1.2 \%$ and $\eta = 0.4 \%$), respectively, which leads to either underestimation or overestimation in the lattice parameter of SnTe compared to the extracted values from the experiment. Regarding this, we also performed calculations with $4.25$ \r{A} from experiment, which stretches the NbSe$_2$ without compromising its electronic properties. In this same new scenario, we also performed additional DFT calculations for x=2 and 3. Figure~\ref{prove_3layer}a shows the relaxed structure when $x=2$, where 2 ALs have strong distortion due to the strain and interaction with the NbSe$_2$ monolayer. At the bottom layer, Sn atoms appear to be much closer to the NbSe$_2$ as opposed to the Te atoms. Interestingly, the same effect at the bottom layer is observed for $x$= 3 and 4, shown in Figs. \ref{prove_3layer}b and c, respectively, while the top few ALs tend to behave like isolated layers in SnTe. Specifically, for $x$=4, the charge difference shown in Fig.~\ref{prove_3layer}c reveals that the bottom 1-AL SnTe bonds to the NbSe$_2$ underneath through the Sn atoms, driving it into a different chemical environment that effectively isolates the remaining 3 atomic layers at the top. The structure shown in the main text is reproduced in Fig.~\ref{prove_3layer}c for comparison . The height profile obtained from our theoretical model agrees well with experimental measurements. In summary, our bilayer SnTe behaves as a bottom 1-AL wetting layer with an effective strained 3-AL system.

\begin{figure*}[h!]
    \centering
    \includegraphics[width = 0.9\textwidth]{SI Figures/prove_3layer.png}
    \caption{Relaxed structures of the SnTe/NbSe$_2$ heterostructures with x= 2, 3 and 4 in \textbf{a}, \textbf{b} and \textbf{c}, respectively. \textbf{c} also shows the charge difference between the heterostructure and the respective isolated counterpart as an isosurface with isovalue 0.003. Light blue represents charge depletion while yellow represents charge accumulation.}
\label{prove_3layer}
\end{figure*}

\section{\textbf{Experimental observation of topological edge states}}

\begin{figure*}[h!]
    \centering
    \includegraphics[width = 0.9\textwidth]{SI Figures/Exclude surface states.png}
    \caption{\textbf{a}, Tunneling spectra taken in the middle of NbSe$_2$, monolayer and bilayer SnTe. \textbf{b}, Tunneling spectra taken in the middle and edge of bilayer SnTe. Long-range spectra in monolayer \textbf{c} and bilayer SnTe \textbf{d}.}
\label{Exclude surface states}
\end{figure*}

\begin{figure*}[h!]
    \centering
    \includegraphics[width =\textwidth]{SI Figures/Two pairs of edge states in isolated edge.png}
    \caption{\textbf{Evolution of edge states.} \textbf{a}, STM image of a bilayer islands on monolayer SnTe (\textit{V$_b$}=0.5 V, \textit{I}=60 pA). \textbf{b}, A series of tunneling spectra were measured from the center of the bilayer to the edge, along the red line shown in Figure~\ref{Edge states at an isolated edge}\textbf{a}. The two green dashed arrows indicate the energy positions of two peaks 1 and 2 associated with edge states. The two purple dashed rectangular frames are highlighted to illustrate the evolution of peaks 1 and 2 from the bulk to the edge. The two dark yellow dashed arrows indicate that the two shoulder peaks of the conduction band gradually shrink to peak 3 from the bulk to the edge. The energy position of the peak 3 is marked by the purple dashed arrow. The \dIdV spectra are shifted equidistantly vertically for clarity. \textbf{c}, Tunneling spectra from the bilayer to the monolayer via the isolated bilayer edge along the linecut in \textbf{a}. Two dashed ovals outline two peaks associated with helical edge states.}
\label{Edge states at an isolated edge}
\end{figure*}

\begin{figure*}[h!]
    \centering
    \includegraphics[width =0.5\textwidth]{SI Figures/Decay of first edge states into bilayer.png}
    \caption{Intensity of the peak 1 decays into the bilayer as a function of distance away from the edge. The behavior is fitted by an exponential function with the decaying length of 0.38 nm. }
\label{Decay of edge states}
\end{figure*}

Figure~\ref{Exclude surface states}a demonstrates the short bias range tunneling spectra taken from the center of the monolayer, bilayer SnTe, and NbSe$_2$ surfaces. Notably, the spectrum on bilayer SnTe is featureless in the bias ranges where the edge states are located as highlighted in Fig.~\ref{Exclude surface states}b. Instead, discrete peaks appear near the bulk gap in the middle of the bilayer, attributed to quantum well states (see discussion below). Figs. \ref{Exclude surface states}c and d present the large bias range \dIdV spectra. We observe that the \dIdV intensity in both monolayer and bilayer SnTe reaches a minimum between 0.5 and 1 V, suggesting a possible bulk band gap. The conduction band in the monolayer exhibits a pronounced peak around $2$ V, while the bilayer shows two shoulders between 1 and 2 V. These features are distinct from those observed in monolayer and bilayer SnTe on graphene\cite{doi:10.1126/science.aad8609}. 

Additionally, we acquired line spectra from the middle of the bilayer towards the monolayer (see Figure~\ref{Edge states at an isolated edge}a). As demonstrated in Fig.~\ref{Edge states at an isolated edge}b, the peak 1 (the first edge state) emerges and gradually increases in intensity as the tip approaches the island edge, as outlined by the purple dashed rectangular frame. In contrast, the two shoulders associated with the conduction band gradually shrink into a single peak 3 (around 1.5 V) from the center of the island to the edge, as indicated by the dark yellow arrows. Meanwhile, an additional peak 2 associated with edge states (the second edge state) emerges at around $1.8$ V at the edge, as indicated by the green dashed arrow. The same data can be plotted in a color scale plot to highlight the spatial localization of the edge state peaks 1 and 2 (Fig.~\ref{Edge states at an isolated edge}c ). The spatial decay of the peak 1 is plotted separately in Fig.~\ref{Decay of edge states}.

\section{\textbf{Determination of bulk gap}}

\begin{figure*}[h!]
    \centering
    \includegraphics[width = 0.5\textwidth]{SI Figures/QuantumWellStates.png}
    \caption{\textbf{Quantum well states in bilayer SnTe.} Tunneling spectra in the monolayer and bilayer SnTe. The bilayer bulk states can change with different-size islands.} 
\label{QuantumWellStates}
\end{figure*}

\begin{figure*}[h!]
    \centering
    \includegraphics[width = \textwidth]{SI Figures/NegativeDifferentialResistance.png}
    \caption{\textbf{a}, Energy diagrams in STM tunneling junctions between degenerate SnTe and the tip. \textbf{b} and \textbf{c} show the experimental (calculated) current-voltage characteristic curve and differential conductance. The calculated current is slightly shifted upwards to coincide with the experimental observations. \textbf{d} and \textbf{e},  Similar curves and conductance over a larger bilayer island. } 
\label{NegativeDifferentialResistance}
\end{figure*}

In the bulk of bilayer SnTe, the \dIdV spectra exhibit a series of discrete peaks, which vary with islands size (see Figure~\ref{QuantumWellStates}). They most likely are related to quantum well states or Fabry-Perot resonances formed in the SnTe wetting layer or in the NbSe2 substrate under the SnTe island\cite{PhysRevLett.108.066809,Herrera2023}. 

Now we focus on determining the bulk gap of bilayer SnTe. Our bilayer SnTe is highly p-type
and exhibits metallic behavior, making it difficult to directly measure the band gap using tunneling differential conductance. However, negative differential resistance (NDR) in \dIdV spectra allows us to estimate the band gap as described below. The tunneling model is demonstrated in Figure~\ref{NegativeDifferentialResistance}a. Electron tunneling takes place between the metallic tip and a degenerate semiconductor SnTe through the vacuum layer. The transmission coefficient $T$ within the Wentzel-Kramers-Brillouin formalism changes with the bias voltage as follows,
\begin{equation}
    T\approx16\frac{E}{V}(1-\frac{E}{V})\exp\frac{-2d\sqrt{2m(V-E)}}{\hbar}
\label{equation1}
\end{equation}
in the limit of ${\exp^{\frac{-d\sqrt{2m(V-E)}}{\hbar}}\gg1}$, where $d$ is the potential barrier width and $V$ is the barrier height. As the bias voltage increases such that the valence band of the sample aligns with the Fermi level of the tip, the tunneling current increases. When the bias voltage is further increased, the total number of allowed empty states in the  degenerate p-type SnTe layer remains unchanged, as electrons cannot tunnel into the band gap. Meanwhile, the tunneling barrier height increases, leading to a reduction in tunneling probability (as reflected in the exponential term in Eq. \ref{equation1}) and, consequently, a decrease in tunneling current. This results in the observation of negative resistance in the current-voltage curve and differential conductance\cite{doi:10.1126/science.183.4130.1149}.

When the bias voltage exceeds the minimum edge of the conduction band, a new tunneling path from the tip to the conduction band of the sample opens, causing the tunneling current to increase again. Based on the above theory, we applied Esaki's equations from Ref. \cite{PhysRevLett.16.1108} to extract the band gap in the bilayer SnTe. Interestingly, the tunneling current reaches a maximum around e${V=E_{F_p}}$, and then decreases with increasing $V$. The current contributed by electrons tunneling into $E_{C_p}$ starts at e${V=E_{F_p}+E_g}$ from the tip to the sample and soon offsets the current from negative resistance with the increasing bias voltage. We fitted the current-voltage characteristics curves and \dIdV spectra and estimated the band gap to be around 0.3 eV spanning from 0.5 to 0.8 eV in the small island, as illustrated in Fig.~\ref{NegativeDifferentialResistance}b and c. Similarly, the bulk gap in the larger island is estimated to be about 0.2 eV extending from 0.5 to 0.7 eV (Fig.~\ref{NegativeDifferentialResistance}d and e).

\section{\textbf{Effect of point defects and external electric field}}

We already discuss the effect of point defects and moiré potentials on the edge states in the experiments in the main text. Here, we complement this by testing these effects theoretically. As shown in Figure~\ref{Robustness}a, we add moiré potential into a pristine TCI for simulating the effect of moiré patterns on mirror symmetry. The result shows that the edge modes are resilient against it and thus edge states remain gapless. In contrast, introduction of point defects at the edge can completely destroy the edge states because the intrinsic atomic defects strongly break the mirror symmetry.

In addition, we test the effect of external electric fields on the edge states both experimentally and computationally. When we push the STM tip closer and closer to the sample, correspondingly, the increasing tunneling current and thus stronger and stronger perpendicular electric field to the edge states, we find the edge states remain all the way (Figure.~\ref{3layerbands with electric field}a). Supposing the tip-sample separation is 0.4 nm at 1100 pA, and the thickness of 3-AL SnTe ultrathin films is about 0.64 nm estimated from the calculations. Then the strength of electric field applied to edge states at 0.45 eV is 0.4 V/nm.  To exclude the screening effect from the metallic NbSe$_2$ substrate, we calculated the effect of a perpendicular electric field on the band inversion of an isolated 3-AL SnTe system with compressive strain. As shown in Fig.~\ref{3layerbands with electric field}b, the results show that even an electric field up to 10.8 V/nm cannot break the band inversion at the $X$ point. This is consistent with our experimental observations, but very different from the past theory where the application of small electric field of 0.03 V/nm is able to break the mirror symmetry\cite{Liu2014}.

\begin{figure*}[h!]
    \centering
    \includegraphics[width = \textwidth]{SI Figures/Simulation of edge states.png}
    \caption{Theoretical simulations of resilience of pristine edge states against moiré potential and point defects at the edge.}
\label{Robustness}
\end{figure*}

\begin{figure*}[h!]
    \centering
    \includegraphics[width = \textwidth]{SI Figures/3layerbands_ef.png}
    \caption{\textbf{Experimental and computational tests of edge states robust against external electric field.} \textbf{a}, Edge states as a function of tunneling current. \textbf{b}, Evolution of band structures of the 3-AL SnTe system with the lattice constant of 4.33 \AA obtained with perpendicular electric fields. Spin-orbit coupling is included in all the calculations. Green and orange colors indicate the projections onto the p orbitals of Sn and Te atoms, respectively.}
\label{3layerbands with electric field}
\end{figure*}

\section{Visualization of topological edge states}

\begin{figure*}[h!]
    \centering
    \includegraphics[width = 0.9\textwidth]{SI Figures/dIdVmaps.png}
    \caption{\textbf{\dIdV maps at different bias voltages.} \textbf{a}, Atomically resolved STM image of the bilayer SnTe islands (\textit{V$_b$}=0.5 V, \textit{I}=240 pA). Constant current \dIdV maps from \textbf{b} to \textbf{f} taken at 0.2 V, 0.4 V, 0.5 V, 0.6 V and 0.8 V, respectively.}
\label{ExtendedData dIdV maps}
\end{figure*}

\begin{figure*}[h!]
    \centering
    \includegraphics[width =\textwidth]{SI Figures/Irregular bilayer.png}
    \caption{\textbf{Edge states in irregular bilayer SnTe.} \textbf{a}, Typical STM image of an irregular bilayer SnTe island (\textit{V$_b$}=0.5 V, \textit{I}=20 pA). \textbf{b}, Tunneling \dIdV spectra in the bulk of regular and irregular bilayer islands. \textbf{c}, Short-range \dIdV spectra taken in the bulk and edge of the irregular bilayer, marked by blue and red points in \textbf{a}.}
\label{Irregular bilayer}
\end{figure*}

\begin{figure*}[h!]
    \centering
    \includegraphics[width = 0.9\textwidth]{SI Figures/Large scale dIdV map.png}
    \caption{\textbf{Overview of topological edge states.} \textbf{a}, Large-scale STM topography for bilayer SnTe (\textit{V$_b$}=0.5 V, \textit{I}=200 pA). 'A', 'B','C','D' indicate four types of edge configurations for two phases in bilayer SnTe. \textbf{b}, Constant current \dIdV map at 0.5 V.}
\label{Large dIdV maps}
\end{figure*}

\begin{figure*}[h!]
    \centering
    \includegraphics[width = 0.85\textwidth]{SI Figures/dif-terminations.png}
    \caption{Band structures of 3-AL SnTe with different atomic terminations.} 
\label{Different terminations}
\end{figure*}

To visualize spatial variation of edge states, we made a series of \dIdV maps at different energies. Figure~\ref{ExtendedData dIdV maps}a demonstrates the topography of two adjacent bilayer islands on the monolayer. As shown in Figs. \ref{ExtendedData dIdV maps}b-f, at 0.5 V and 0.6 V, the edge states are conducting along the sharp edges within the bulk gap. At 0.2 V and 0.8 V, the edge states fade away and disappear. Meanwhile, SnTe has another phase clearly distinguished by the topography. As shown in Figure~\ref{Irregular bilayer}a, this phase features an irregular shape and also follows the monolayer moiré patterns. Its conduction band exhibits a sharp peak around 1.5 V (Fig.~\ref{Irregular bilayer}b), quite different from the regular bilayer. At edges, we can also observe an edge state peak within nearly insulating bulk states around 0.5 V in Fig.~\ref{Irregular bilayer}c. Its lattice constants are estimated to be 4.31 \AA. From our theoretical calculations in Figure~2 of the main text, the irregular bilayer should be a 2D TCI, too.  

To determine what edge configurations can host edge states, we performed a large-scale scan over bilayer islands, as shown in Figure~\ref{Large dIdV maps}a. Regular bilayer islands manifest a rectangular shape with sharp and straight edges, while irregular bilayer islands have multiple rough and winding edges. Four typical edge configurations are indicated in the two SnTe phases. Then we took a \dIdV map at 0.5 V to visualize the edge states, as shown in Fig.~\ref{Large dIdV maps}b. We find that only along the sharp edges with specific atomic terminations like the edges 'A' and 'D' in the two phases, there are conducting edge states. Otherwise, the edge states are completely absent. This can be compared with our calculations (Figure~\ref{Different terminations}) that show how the band structures behave for different atomic edge terminations. The band structures are obtained without applying the scissor operator. The results show that only (100) ribbons possess edge states and the atomic terminations are either Sn or Te atoms.

\begin{figure*}[h!]
    \centering
    \includegraphics[width = 0.95\textwidth]{SI Figures/Conduction band for bulk.png}
    \caption{\textbf{a}, Conduction band in bilayer bulk. \textbf{b} and \textbf{c}, Conduction band along (across) stripes in bilayer islands. The red rectangular frames indicate the edge states and the green frames show a band bending close to edge. }
\label{Bulk conduction band}
\end{figure*}

From the strain map, we know the bright stripes host smaller compressive strain while the dark ones harbours larger strain. The strain has impact on the conduction band, as shown in Figure~\ref{Bulk conduction band}. The conduction band in bright stripes shifts towards a lower energy than that of dark stripes. The conduction band along and across the stripes modulates periodically with the strain field. Conduction bands at the opposite ends of the same stripe both bend downwards. 

%\begin{figure*}[h!]
    %%%% \centering
     %%%\includegraphics[width = 0.5\textwidth]{SI Figures/Anticorrelation spectra.png}
    %% \caption{\dIdV spectra taken at the different sites at the edge. The two edge state peaks are anticorrelated with the strain field as marked by the green arrows.}
     %\label{Anticorrelation}
%\end{figure*}

\section{Quantum tunneling between adjacent edge states}
\begin{figure*}[h!]
     \centering
     \includegraphics[width =\textwidth]{SI Figures/Tunneling model.png}
     \caption{\textbf{a}, Schematic representation of adjacent edge channels from two bilayer islands. One pair of edge states is indicated by red lines and the other one marked by green lines. The proximity effect is accounted for by an effective barrier potential \textit{V} between the two islands. \textbf{b}, The Dirac points at $X$ shifts towards lower energy with the decreasing $\textit{V}$, as indicated by black arrows. \textbf{c}, The calculated energy shift of the low-energy Dirac point can be fitted by an exponential function of $\lambda d$, where $\lambda d =\ln\left(\frac{V}{t_0}\right)$, $t_0$ is the hopping energy of SnTe orbitals (on the order of eV), $\lambda$ is a finite phenomenological parameter that can be adjusted based on experimental data and $d$ is the distance between adjacent edges. }
     \label{Tuneling coupling}
\end{figure*}

\newpage

%\bibliographystyle{naturemag1}
%\bibliography{biblio}
\bibliography{Ref}